\documentclass[american,superscriptaddress,aps,reprint,prl]{revtex4-2}
\usepackage[T1]{fontenc}
\usepackage[utf8]{inputenc}
\usepackage{bm}
\usepackage{amsmath}
\usepackage{amssymb}
\usepackage{comment}
\usepackage{graphicx} % For including images
\usepackage{etoolbox}
\usepackage{xcolor}

\apptocmd{\thebibliography}{\sloppy}{}{}
\usepackage[utf8]{inputenc}

\newcommand{\rr}{\boldsymbol{\rho}}
\newcommand{\rravg}{\rr}
\newcommand{\rrI}{\rr^\text{I}}
\newcommand{\rrII}{\rr^\text{II}}

\makeatletter

\usepackage{newtxtext}

\makeatother

\usepackage{babel}
\begin{document}
\title{Spatiotemporal organization of chemical oscillators via phase separation}
\author{Jonathan Bauermann}
\email{jbauermann@fas.harvard.edu}
\affiliation{Department of Physics, Harvard University, Cambridge, MA 02138, USA}
\author{Giacomo Bartolucci}
\affiliation{Department of Physics, Universitat de Barcelona, 08007  Barcelona, Spain}
\author{Artemy Kolchinsky}
\affiliation{ICREA-Complex Systems Lab, Universitat Pompeu Fabra, 08003 Barcelona, Spain}

\begin{abstract}
We study chemical oscillators in the presence of phase separation. 
By imposing timescale separation between slow reactions and fast diffusion, we define a dynamics at phase equilibrium for the relevant degrees of freedom.
We demonstrate that phase separation affects reaction kinetics by localizing reactants within phases, allowing for 
control of oscillator frequency and amplitude. 
The analysis is validated with a spatial model. Finally, relaxing the timescale separation between reactions and diffusion leads to waves of phase equilibria at mesoscopic scales.

\end{abstract}
\maketitle

\textit{Introduction.--- } 
The discovery of biomolecular condensates in the  cytoplasm~\cite{brangwynneGermlineGranulesAre2009}, believed to form via phase separation,  has led to increasing interest in the role of phase separation in cell biology~\cite{Banani2017, Boeynaems2018, Hyman2014, Mitrea2016}. 
Biomolecular condensates have been associated with a variety of cellular functions, including compartmentalisation of chemical reactions in space~\cite{Strulson2012, Nakashima2019, Weber2019}.
Phase separation, together with reactions, may have also played an important role in abiogenesis. In fact, more than a century ago, Oparin and Haldane speculated that chemically active droplets might have been the first ``metabolic units'' at the origin of life~\cite{Oparin1953, Haldane1929}.
Nowadays, phase separation is also drawing attention in ecology as a potential mechanism for the formation of patchy ecosystems~\cite{liu2013phase, siteur2023phase}.

In the context of nonequilibrium chemical reactions, phase separation can influence reaction kinetics by partitioning space and affecting local concentrations of reactants. Conversely, reactions can control the emergence of phases by altering chemical concentrations. As a result, highly nontrivial phenomena can result from coupling phase separation to nonequilibrium reactions, including pattern formation~\cite{aslyamov2023nonideal, avanzini2024nonequilibrium} and nonequilibrium effects on droplet composition~\cite{kirschbaumControllingBiomolecularCondensates2021,Cho2023, Laha2024}, ripening~\cite{Glotzer1995, Zwicker2015, Wurtz2018, Kumar2023, bauermann2025sizedist}, nucleation~\cite{Ziethen2023, Cho2023b}, division~\cite{zwickerGrowthDivisionActive2017, Bauermann2022}, and propulsion~\cite{Demarchi2023, Hfner2024, Goychuk2025}.

Recent work has also suggested that phase separation may play an important role in various oscillatory biomolecular processes, including DNA repair~\cite{heltberg2022enhanced} and circadian clocks~\cite{zhuang2023circadian,tariq2024phosphorylation}. However, while it has been shown that phase separation can trigger oscillations~\cite{Haugerud2025, Sastre2025}, little is known about the general relationship between phase separation and the temporal organization of oscillatory reaction networks~\cite{Smokers2024}.
To close this gap, we analyze the effects of phase separation in a simple model of a nonlinear chemical oscillator, where the co-localization of chemical species via phase separation allows for targeted control of oscillator frequency and amplitude. 
We perform this analysis by considering a dynamics at phase equilibrium.
Our techniques can be straightforwardly generalized to more coexisting phases and other nonlinear reaction networks, thus paving the way for studying the effects of physical interactions on nonequilibrium dynamics in a new light.

\textit{Chemical Oscillator.--- } 
We consider a simple chemical oscillator with three species $A,B,C$ and three irreversible reactions:
\begin{equation}
    A+B \xrightarrow{k_{AB}} 2 A,\;\;\; B+C \xrightarrow{k_{BC}} 2 B , \;\;\; C+A \xrightarrow{k_{CA}} 2 C , \;
\end{equation}
with second-order rate constants $k_{ij}$, known as the ``rock-paper-scissors'' (RPS) scheme~\cite{Smith2012-py,Sinervo1996}. 

Before proceeding to study phase separation, we consider the case of a well-mixed ideal system characterized by average (volume) concentrations
$\rr=(\rho_A,\rho_B,\rho_C)$. Assuming  mass-action kinetics, the deterministic dynamics of these concentrations read
\begin{align}
    \frac{d\rr}{dt} = \mathcal{R}(\rr), \quad \mathcal{R}(\rr)= \begin{bmatrix}
    k_{AB} \rho_A \rho_B  - k_{CA} \rho_C \rho_A \\
    k_{BC} \rho_B \rho_C  - k_{AB} \rho_A \rho_B\\
    k_{CA} \rho_C \rho_A  - k_{BC} \rho_B \rho_C \end{bmatrix} \;.
    \label{eq:reac_flux}
\end{align}
These dynamics involve two degrees of freedom, because the total concentration $\psi=\rho_A + \rho_B + \rho_C$ is a conserved quantity determined by initial conditions.   In the following, we set the rate constants as $k_{AB}=k_{BC}=k_{CA}=1$.

\begin{figure}[!t] 
    \includegraphics[width=0.96\linewidth]{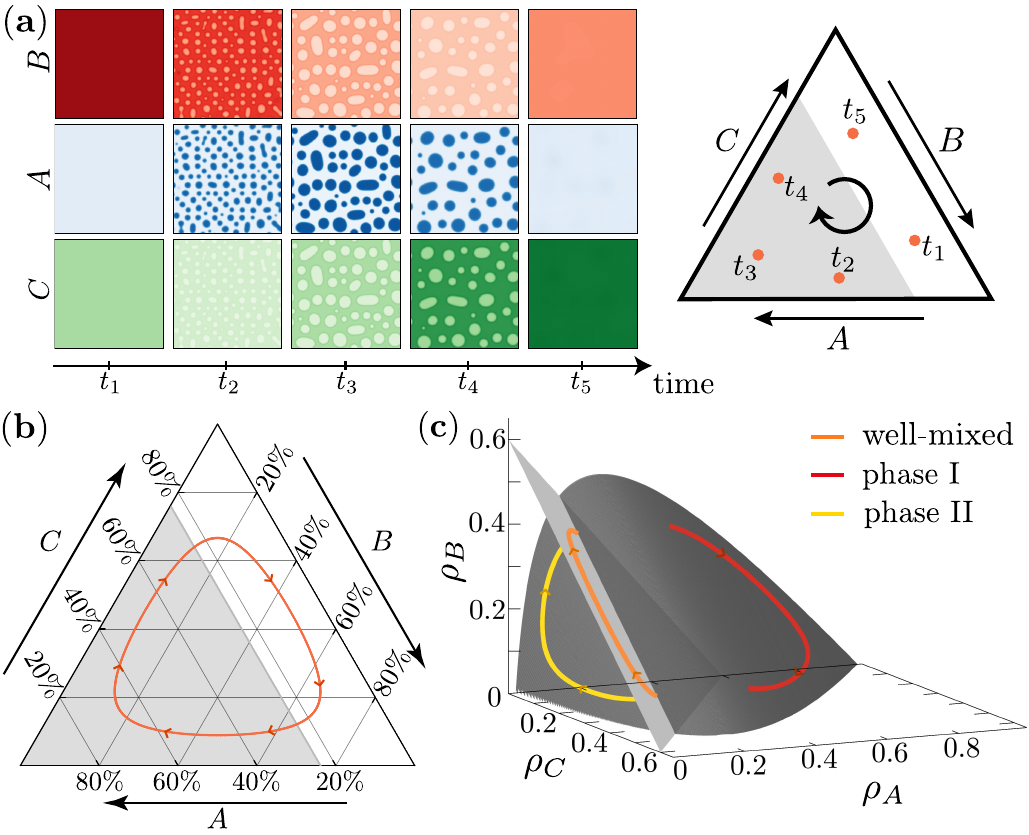}
    \caption{
    \textbf{Chemical oscillator in the presence of coexisting phases.}
    (a) Spatial profiles (left) of $\rr$ and corresponding spatial averages sketched in concentration space (right) at five time points in a spatial model.
    (b) Trajectory of average (volume) concentrations in $\boldsymbol{\rho}$ in concentration space  (orange).
    Within the binodal domain (grey shaded), phases coexist at equilibrium; outside the binodal domain (white), equilibrium is well-mixed. 
    (c) Trajectories of $\rravg$ (orange) outside the binodal, and phase concentrations $\rrI$ (red) and $\rrII$ (yellow) at the binodal. Dark grey: binodal manifold;
    light grey: plane of constant total concentration $\psi$.
    Parameters: (a) $\chi_{AS} = 4$, $\chi_{AB} = \chi_{AC} = 1$, see Fig.~\ref{fig:spatial_slow} for further details.
    (b,c) $\psi = 0.6$, $\chi_{AS} = 4$, $\chi_{AB} = \chi_{AC} = 0$.
    }
    \label{fig:phase_space}  
\end{figure}

Every initial condition leads to an oscillation with a fixed amplitude that comes back to itself. 
As we discuss below, this oscillatory trajectory is identical to the average concentration trajectory shown in  Fig.~\ref{fig:phase_space}(b).

\textit{Phase separation.--- } 
To introduce phase separation, we consider interactions between components. We use the Flory-Huggins free energy density, written in dimensionless units as 
\begin{equation}
\label{eq:free_ener}
 f(\rr) = \sum_{i=S,A,B,C} \rho_i \ln\rho_i + \sum_{j=S,B,C} \chi_{Aj} \rho_A \rho_j\,,
\end{equation}
where $S$ is the solvent with concentration $\rho_S= 1-\psi$. The first sum describes the entropic contribution of mixing, while the second sum describes energetic interactions between components that can lead to phase separation.
In this work, we focus on cases where one reactant, we choose $A$, can phase separate, while the others only co-localize in phases. Thus, we only consider interactions between $A$ and components $j=S,B,C$, parameterized by interaction energies  $\chi_{Aj}$.

Reaction dynamics in phase-separating systems are usually studied in spatial models with local equilibrium~\cite{Weber2019}, where each location $\bm{x}$ in volume $V$ is associated with a local concentration vector $\rr(\bm{x})$.
The total free energy of a system, $F=\int_V d\bm{x} \left[f + \sum_i \kappa_i(\nabla \rho_i)^2/2\right]$, is an integral over the local free energy density $f$ and the free energy contribution of spatial concentration gradients, parameterized by $\kappa_i$ for species $i$. 
Spatial gradients of the local chemical potentials $\mu_i(\bm{x}) = \delta F/\delta \rho_i(\bm{x})$ drive diffusive fluxes. The deterministic dynamics of components $i=A,B,C$ then read~\cite{Onsager1931, De_Groot2003}
\begin{equation}
\label{eq:spatial}
    {\partial_t} \rr(\bm{x})  = \nabla \cdot (\bm{\Gamma}(\bm{x}) \nabla \bm{\mu}(\bm{x}) ) 
    + \mathcal{R}(\rr(\bm{x}))
    \, .
\end{equation} 
Here, we choose $\Gamma_{ij} = \gamma \rho_i (\delta_{ij} - \rho_j)$ as the mobility matrix with mobility coefficient $\gamma$, such that the diffusion term in Eq.~\eqref{eq:spatial} reduces to Fick's law when interactions $\chi_{Ai}$ and gradient terms $\kappa_i$ vanish~\cite{Kramer1984, Bo2021}. In this case Eq.~\eqref{eq:spatial} reduces to the classical reaction-diffusion equation for the RPS dynamics~\cite{Reichenbach2007, Reichenbach2008, He2010}.

The spatial model can still exhibit oscillations as a result of the RPS reaction term. In addition, the concentration fields may exhibit phase separation when the interactions in Eq.~\eqref{eq:free_ener} are sufficiently strong.  
As detailed in the Appendix, we choose $\chi_{AS}$ to be sufficiently large so that two phases form: an $A$-rich phase I and an $A$-poor phase II. We set $\chi_{AB}$ and $\chi_{AC}$ sufficiently small so that $B$ and $C$ only localize inside/outside the two phases, but do not give rise to additional phases. 
Depending on whether $\chi_{AB}< 0$, $\chi_{AB}= 0$, or  $\chi_{AB}> 0$, the concentration of $B$ is higher, equal, or lower (respectively) in phase I, and similarly for $\chi_{AC}$ and $C$.

As an illustration, in Fig.~\ref{fig:phase_space}(a), we show the spatial concentration profiles of $A$, $B$, and $C$ from a typical trajectory of the spatial model in two dimensions, assuming fast diffusion and slow reactions. We show five time points within one period of an oscillation (left), illustrating nucleation, growth, shrinking, and dissolution of droplets. Furthermore, we sketch the spatial averaged concentrations at these time points in the concentration space (right), showing how the system cycles through the region of high $A$ concentrations where droplets can form (shaded in grey).

The previous example illustrates how reaction dynamics in phase-separating systems can be solved numerically using the spatial model.
However, this approach is computationally intensive and introduces complexities, such as interface effects and nucleation barriers. In particular, as explained below, homogeneous solutions are metastable for some concentrations, and fluctuations in Eq.~\eqref{eq:spatial} are required to nucleate droplets. Furthermore, the spatial model is difficult to analyze, for instance, to study stability analysis at phase equilibrium.

To circumvent these difficulties, in the following we consider reaction dynamics at global phase equilibrium. 
This leads to a low-dimensional system that is numerically and analytically tractable, allowing us to gain insight on the control of chemical oscillations through phase separation.
We compare the results between the coarse-grained model and the spatial models in the Appendix and, at the end of this paper, we examine  cases where diffusion and reactions occur on similar timescales.

\textit{Reaction dynamics at phase equilibrium.--- } 
Ideal systems are guaranteed to be well-mixed when diffusion is fast compared to reactions. Similarly, here, we consider non-ideal systems, where fast diffusion establishes a global phase equilibrium instantaneously relative to the reaction timescale. 
For a given average concentration $\rr$, phase separation is favored 
whenever there exist two phases, characterized by concentrations $\rr^\text{I/II}$ and relative volume  $v\in [0,1]$ of phase I, such that an averaged free energy
\begin{equation}
\label{eq:favg}
    f^\text{avg}(v,\rrI,\rrII) = v f(\rrI) + (1-v) f(\rrII)\; ,
\end{equation}
subject to the constraint
\begin{equation}
   \rravg = v \rrI + (1-v)\rrII \;,
   \label{eq:avg_con}
\end{equation}
exits with $f^\text{avg}(v,\rrI,\rrII)<f(\rravg)$.
The boundary of the domain of $\rr$ for which this condition is satisfied is called the binodal. Within this region, the seven equilibrium values of $(v,\rrI,\rrII)$ are found as the global minimum of $f^\text{avg}$ under the constraint~\eqref{eq:avg_con}~\cite{Safran2019, Kardar2007}. 

Within the binodal, reactions occur in each phase, so the average concentration $\rr$ evolves as~\cite{bauermann2022chemical}
\begin{align}
    \frac{d\rravg(v,\rrI,\rrII)}{dt} = v \mathcal{R}(\rrI)+(1-v)\mathcal{R}(\rrII)\,.
    \label{eq:dynps}
\end{align}
As in the ideal system, the total average concentration $\psi$ is conserved; thus, the dynamics of $\rr$ is still two-dimensional. 

During the temporal evolution, the system can move in and out of the binodal domain. 
Thus, the full dynamics is a combination of Eq.~\eqref{eq:reac_flux} and Eq.~\eqref{eq:dynps}. Whenever the system is outside the binodal domain, we use Eq.~\eqref{eq:reac_flux}. However, when the state variables enter the binodal domain, we use Eq.~\eqref{eq:dynps}.

In Fig~\ref{fig:phase_space}(b,c), we illustrate chemical oscillations in the special case $\chi_{AB}=\chi_{AC}=0$. For these parameters,  $B$ and $C$ do not localize preferentially in either phase, and remain homogeneously distributed. 
Because the RPS fluxes are linear in the concentration of each individual reactant, the dynamics in Eq.~\eqref{eq:dynps} differ from Eq.~\eqref{eq:reac_flux} only when two or more reactants (not only $A$) have unequal concentrations between the phases. 
Therefore, when $\chi_{AB}=\chi_{AC}=0$, the average concentrations $\rravg$ undergoes the same trajectory for all values of $\chi_{AS}$, shown in Fig.~\ref{fig:phase_space}(b).
In Fig.~\ref{fig:phase_space}(c), we show the dynamics in the three-dimensional phase space of the concentrations $\rr$.
Outside of the binodal domain (dark grey), the trajectory (orange line) lies within the plane of constant $\psi$ (light grey). However, when the oscillation enters the binodal domain, the trajectories of $\rrI$ (red) and $\rrII$ (yellow) separate and run along the binodal manifold.

\begin{figure}[!t] 
    \includegraphics[width=0.96\linewidth]{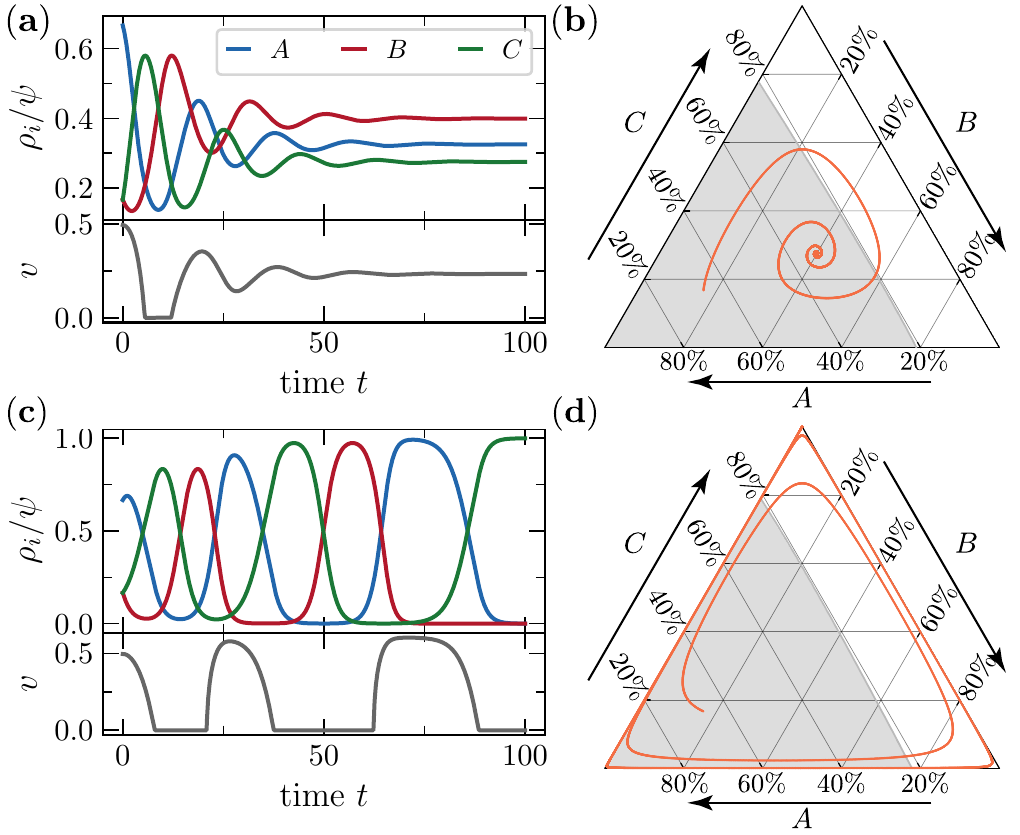}
    \caption{ \textbf{Phase separation controls oscillations.}\label{fig:psRPS} 
    Left: oscillations of the average concentrations $\rravg$ and volume $v$. Right: trajectories of the average concentrations, with region within binodal domain in grey. (a,b) For $\chi_{AB}=1, \chi_{AC}=-1$, oscillations are damped. (c,d) For $\chi_{AB}=-1, \chi_{AC}=1$, oscillation are enhanced.  $\psi = 0.6, \chi_{AS} = 4$. 
    } 
\end{figure}

\textit{Phase separation controls the oscillations.--- } 
Phase separation can localize $B$ and $C$ in the $A$-rich/$A$-poor phases. In turn, this can accelerate or decelerate reactions in the corresponding phases, resulting in changes in oscillator amplitude and frequency. 
The fact that oscillator amplitude and frequency can be controlled via phase separation is a key finding of our work. 
In the following section, we study this phenomenon quantitatively and discuss the underlying physical mechanism. 

We illustrate the interplay of phase separation and oscillations for two different values of $\chi_{AB},\chi_{AC}$ in Fig.~\ref{fig:psRPS}.
For a system where $B$ localizes in the $A$-rich phase while $C$ localizes in the $A$-poor phase ($\chi_{AB}>0,\,\chi_{AC}<0$), the oscillation is damped, Fig.~\ref{fig:psRPS}(a,b). 
For the chosen initial conditions, the average concentration $\rr$ (orange) only leaves the binodal domain (grey in Fig.~\ref{fig:psRPS}(b)) in the first oscillation, leading to a transient where the volume of the $A$-rich phase $v$ vanishes, bottom of Fig.~\ref{fig:psRPS}(a).  
Afterwards, the dynamics settles at a stationary state with coexisting phases ($0 < v < 1$).
However, when $B$ localizes in $A$-rich phase and $C$ in the $A$-poor one ($\chi_{AB}<0,\,\chi_{AC}>0$), the oscillation amplitude increases over time, and there are longer and longer intervals in which the mixture is composed mainly of a single solute, i.e. $\psi \sim \rho_i$.
When this solute is $A$, two phases coexist; otherwise, the system is well-mixed (see Fig.~\ref{fig:psRPS}(c,d)).

\begin{figure}[!t] 
    \includegraphics[width=0.96\linewidth]{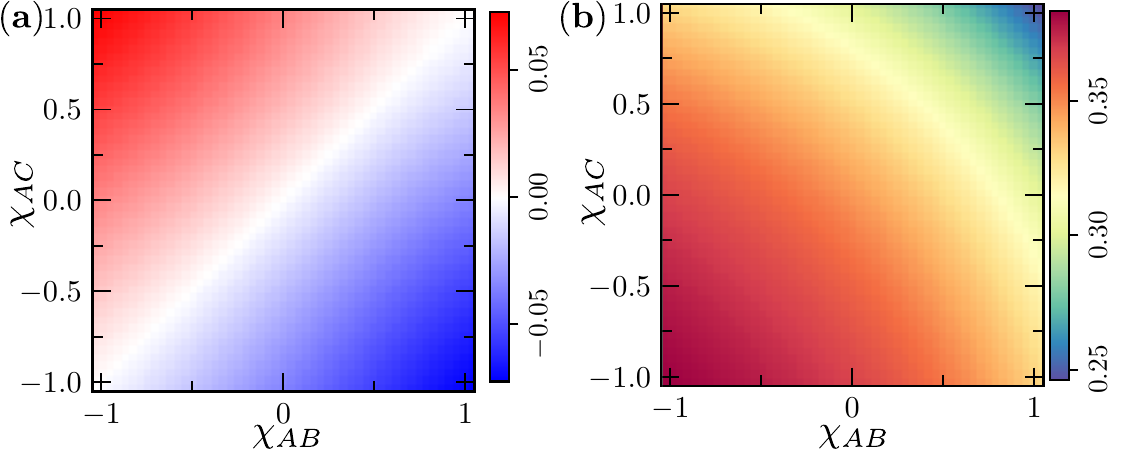}
    \caption{\textbf{Phase separation affects stability and oscillation frequency} at the fixed point. 
    (a) Real and (b) Imaginary part of the relaxation rates of perturbations to the fixed point of Eq.~\eqref{eq:dynps} as a function of $\chi_{AB}$ and $\chi_{AC}$. $\psi = 0.6$, $\chi_{AS} = 4$;
    \label{fig:LSA} 
    } 
\end{figure}

To quantify the effects of localization on oscillator properties, we compute the linear stability of the fixed point of the dynamics. 
Here we focus on parameter regimes for which the fixed point is located within the binodal domain, i.e., involves phase separation. 

Because $\psi$ is a conserved quantity, the fixed point with coexisting phases may be specified by two degrees of freedom, chosen as  $\rr^* = (\rho_A^*,\rho_B^*)$. We perturb this fixed point by the vector $(\delta\rho_A,\delta \rho_B)$.  To first order, the perturbation evolves as
\begin{gather}
    \frac{d}{dt}\begin{bmatrix}
   \delta \rho_A\\
   \delta\rho_B
    \end{bmatrix} = \bm{M} \begin{bmatrix}
   \delta \rho_A\\
   \delta \rho_B
    \end{bmatrix} \,, \\
    M_{ij} =
        \left. \frac{\partial (d \rho_i/dt)}{\partial \rho_j } \right\vert_{\rr^*} = 
        \sum_{n=1,..,7} 
        \left. 
        \frac{\partial (d \rho_i/dt)}{\partial \mathcal{E}_n } 
        \frac{\partial \mathcal{E}_n }{\partial \rho_j} 
        \right\vert_{\rr^*}\;,
        \label{eq:pert0}
\end{gather}
with $i,j = A,B$, and the dummy variables $\mathcal{E}_{1,...,7}=v,\rho^\text{I}_A,\rho^\text{I}_B,\rho^\text{I}_C,\rho^\text{II}_A,\rho^\text{II}_B,\rho^\text{II}_C$. The latter are needed because the dynamics Eq.~\eqref{eq:dynps} depends on $\rravg$ via values of $(v,\rrI,\rrII)$ that are found by minimizing $f^\text{avg}$. 
The derivatives $d\mathcal{E}_n / d \rho_j$ are derived in the Appendix.

The stability of the fixed point $\boldsymbol{\rho}^*$ is given by the two eigenvalues of the $\boldsymbol{M}$~\cite{Strogatz2018} which have the form $\lambda_{1/2} = \lambda_\text{Re} \pm i \lambda_\text{Im}$. 
Note that, in general, finding the fixed point of Eq.~\eqref{eq:dynps} and its corresponding phase equilibrium must be done numerically.
Once found, however, the eigenvalues $\lambda_{1/2}$ of $\boldsymbol{M}$ can be computed in closed form.

In Fig.~\ref{fig:LSA}, we show the real and imaginary parts of these eigenvalues $\lambda_\text{Re}$ and $\lambda_\text{Im}$ in the space of $\chi_{AB}$ and $\chi_{AC}$. 
The sign of $\lambda_\text{Re}$ predicts that perturbations grow when $\chi_{AB}<\chi_{AC}$, are stable for $\chi_{AB}=\chi_{AC}$ and are damped when $\chi_{AB}>\chi_{AC}$.
The imaginary part controls the frequency of the oscillations. The lower the values $\chi_{AB}$ and $\chi_{AC}$, the faster the system oscillates.
These findings are consistent with the examples shown in Fig.~\ref{fig:psRPS}.

The influence of localization on the frequency and amplitude can be explained as follows: if $B$ localizes in the $A$-rich phases ($\chi_{AB}<0$), the production of $A$ via the reaction $A+B \rightarrow 2 A$ is accelerated. 
Similarly, localization of $C$ in the $A$-rich phase ($\chi_{AC}<0$) accelerates the decay of $A$ via the reaction $C+A \rightarrow 2 C$. 
Vice versa, both processes decelerate when the corresponding $\chi_{Aj}>0$.
The third reaction $B+C \rightarrow 2 B$ is unaffected by localization, because it occurs mainly when the average $A$ concentration is low, i.e., when the system is well-mixed.
Thus, as both interaction energies $\chi_{AB},\chi_{AC}$ become negative, the production and the decay of $A$ are accelerated, leading to a faster oscillation; as they become positive, the frequency decreases. 
However, oscillation amplitude is governed by which of the interaction energies is larger. 
For $\chi_{AB} > \chi_{AC}$, the production of $A$ is accelerated more than its decay. Thus, the amplitude increases over time, pushing the system closer and closer to the boundaries of the concentration space.
For $\chi_{AB} < \chi_{AC}$, the oscillation is damped, causing the system to decay toward a fixed point.

\begin{figure}[!t] 
\includegraphics[width=0.96\linewidth]{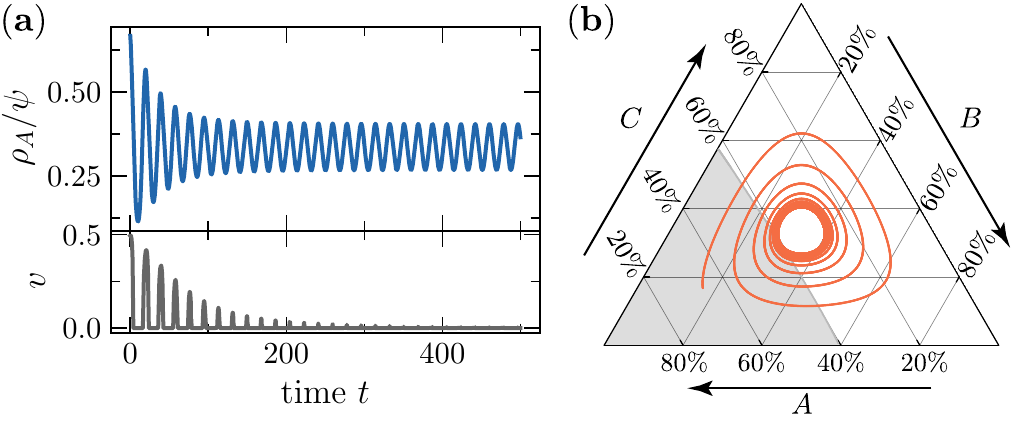}
    \caption{\textbf{Tuning of reaction amplitude} when the fixed point lies in the homogeneous domain:
     (a) Oscillations of the average concentrations $\rho_A$ and relative volume $v$; (b) average concentration trajectory.
     $\psi = 0.6$, $\chi_{AS} = 3$, $\chi_{AB} = 1$, $\chi_{AC} = -1$
    \label{fig:fixpoint_wellmixed} 
    } 
\end{figure}

\textit{Fixed point outside the binodal domain.--- } 
In the previous stability analysis, we focused on situations where the fixed point of the dynamics lies within the binodal domain. However, for lower values of $\chi_{AS}$ or different values of $\psi$, the fixed point may lie outside of the binodal, in the region without phase separation. 
When the initial amplitude is large, such systems still undergo phase separation initially.
When phase separation amplifies the oscillation amplitude ($\chi_{AC}>\chi_{AB}$), the system keeps on entering the binodal domain. 
However, for cases of damped oscillation ($\chi_{AB}>\chi_{AC}$), the amplitude decreases until the system oscillates without entering the binodal domain.
We illustrate this in Fig.~\ref{fig:fixpoint_wellmixed}.
For early times, the trajectory runs through the binodal domain, where the oscillation is dampened, such that the relative volume $v$ gets smaller in each period, as shown in Fig.~\ref{fig:fixpoint_wellmixed}(a). 
Eventually, the oscillation amplitude is sufficiently small such that the oscillation settles into a limit cycle outside the binodal line, see Fig.~\ref{fig:fixpoint_wellmixed}(b).
This mechanism shows that phase separation can not only buffer concentrations at steady state~\cite{deviriPhysicalTheoryBiological2021, Zechner2025}, as demonstrated for biomolecular condensates in the cytosol~\cite{Klosin2020}, but can also buffer the amplitude of chemical oscillations.

\textit{Dynamics with metastability.--- } 
For concentrations $\rravg$ in the outermost part of the binodal, also known as the ``nucleation and growth'' domain, the system can be metastable, and rare macroscopic fluctuations are required to form a droplet above a certain critical size. 
On the other hand, within the ``spinodal'' domain, the well-mixed state is locally unstable and phase separation happens essentially instantaneously~\cite{Bray1994}.
To incorporate this metastability, we accompany the two-dimensional dynamics with a third variable that tracks phase hysteresis induced by metastability.
This variable enforces a switch from Eq.~\eqref{eq:reac_flux} to Eq.~\eqref{eq:dynps} whenever the system is well-mixed and enters the spinodal domain, and a switch from Eq.~\eqref{eq:dynps} to Eq.~\eqref{eq:reac_flux} whenever the system is phase-separated and leaves the binodal domain.

\begin{figure}[!t] 
    \includegraphics[width=0.96\linewidth]{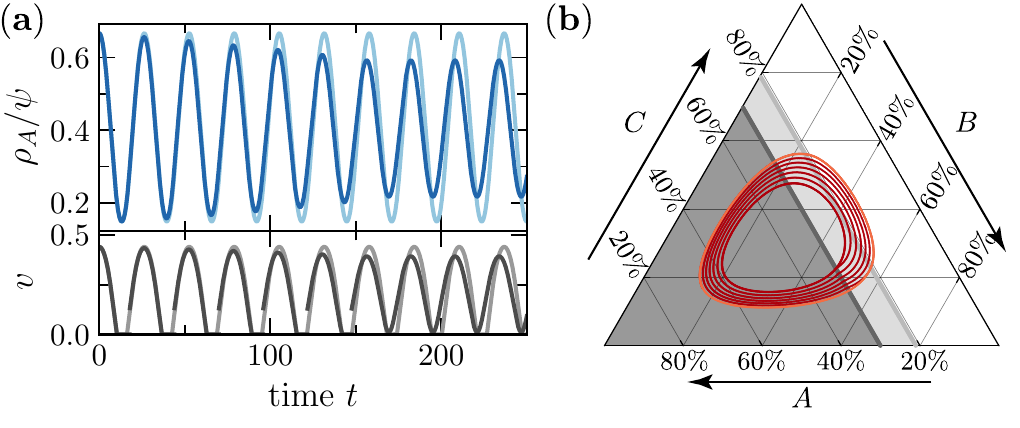}
    \caption{\textbf{Metastability affects oscillations}. \label{fig:spin} 
    Without metastability, phases form as soon as the binodal line is crossed (light colors); with metastability, phases form only once the spinodal (darker colors) are crossed. (a) Oscillations of the average concentration $\rho_A$ and volume $v$. (b) Trajectory of the average concentrations for dynamics without (orange) and with (brown) metastability. 
    $\psi = 0.6$, $\chi_{AS} = 4$, $\chi_{AB} = 1$, $\chi_{AC} = 1$.
    } 
\end{figure}

The presence of metastability affects oscillator dynamics. 
We illustrate this in Fig.~\ref{fig:spin}, where we have chosen $\chi_{AB}=\chi_{AC}=1$. When phases are formed at the binodal oscillations have constant amplitude ($\rho_A$ shown in light blue in Fig.~\ref{fig:spin}(a)) and a closed trajectory (orange in  Fig.~\ref{fig:spin}(b)). 
In contrast, metastability leads to a damped oscillation ($\rho_A$ shown in dark blue in Fig.~\ref{fig:spin}(a)), until the system reaches a limit cycle completely within the binodal domain such that there are always coexisting phases (dark orange in  Fig.~\ref{fig:spin}(b)). 
Production of $A$ is accelerated only when the trajectory is within the spinodal domain (Fig.~\ref{fig:spin}(b), dark grey), while the the decay of $A$ is accelerated as long as the system is within the binodal domain (Fig.~\ref{fig:spin}(b), light grey). 
On the other hand, for $\chi_{AB}=\chi_{AC}<0$, metastability leads to amplification, rather than decay, of oscillations.

The two different dynamics considered above --- one without metastability, where phases form at the binodal, the other with metastability, where phases form at the spinodal --- 
represent two extreme cases. In real systems with noise, phases will form somewhere between the binodal and spinodal lines. 

\begin{figure}[!t] 
    \includegraphics[width=0.96\linewidth]{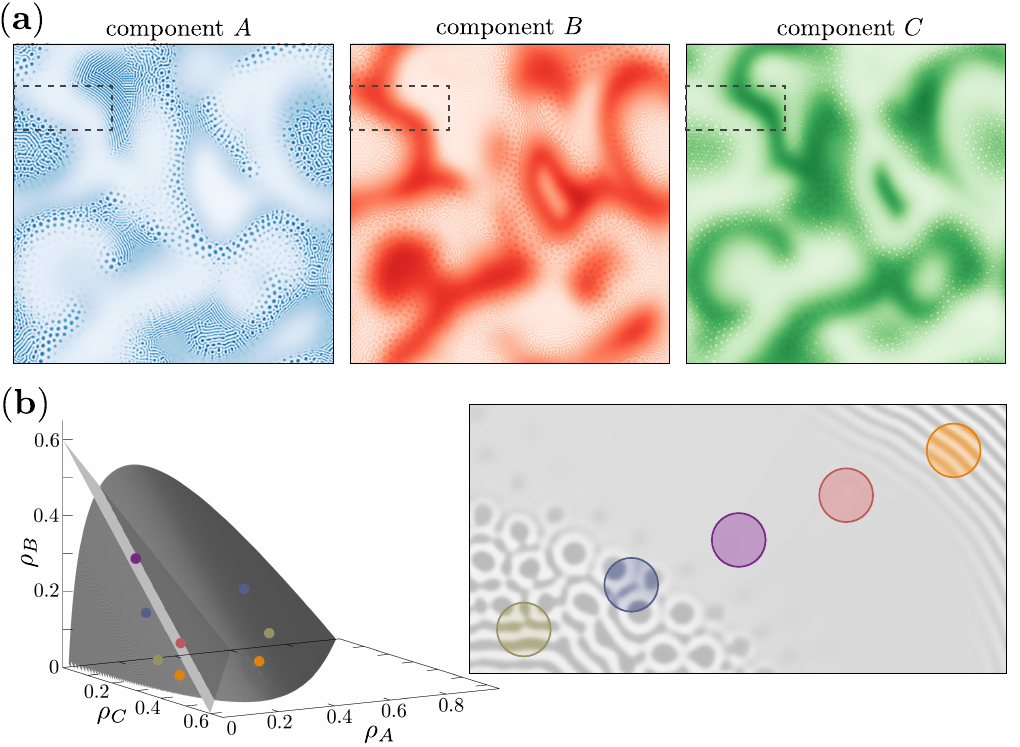}
    \caption{\textbf{Mesoscopic waves of phase equilibria emerge}
    \label{fig:fast_rates}
    in the fast reaction regime.
    (a) Spatial concentrations in a large system.
    (b) Right: local total concentration $\psi$ in a zoomed-in region of space (grey dashed box in top plots). 
    Average concentrations in five mesoscopic environments (colored circles on the right) are shown in concentration space (left), together with the binodal (dark grey) and the plane of conserved $\psi$ (light grey).
    $\psi = 0.6$, $\chi_{AS} = 4$, $\chi_{AB}=-1$, $\chi_{AC}=1$; see Appendix for details.
    } 
\end{figure}

\textit{Fast reactions.--- } 
In our final analysis, we study the case where reaction and diffusion have comparable timescales, so that the system cannot be assumed to be in global phase equilibrium.
We consider the spatial model~\eqref{eq:spatial}, initialized with random concentration fields such that locally the oscillations are at different phases (see Appendix). Whenever the average concentration of $\rho_A$ is high in a mesoscopic region, droplets form in this domain.

For interaction parameters $\chi_{AB}>\chi_{AC}$, the co-localization of $B$ and $C$ into these droplets dampens the oscillations. Thus, on this mesoscopic scale, a stationary state is reached where diffusive fluxes between phases counterbalance the reaction fluxes between components, corresponding to the attracting fixed point of Eq.~\eqref{eq:dynps}. Eventually, the same stationary state is reached everywhere in the system, corresponding to an effective global phase equilibrium.

However, for interactions parameters $\chi_{AB}<\chi_{AC}$, the co-localization of $B$ and $C$ into droplets amplifies oscillations. Here, we find traveling waves and spirals in the concentration fields, 
similar to the case of RPS reaction-diffusion systems without phase separation~\cite{Reichenbach2007,Reichenbach2008, He2010}.
However, at the frontier of $A$-rich waves, droplets are nucleated locally, grow and further amplify the reaction, until they dissolve as the wave travels further. This is illustrated in Fig.~\ref{fig:fast_rates}(a),  where we show concentration fields at one time (see also the movie ``SI-mov-6'' in the SM~\cite{sm}).

In Fig.~\ref{fig:fast_rates}(b), we illustrate how different phase equilibria are selected within these waves. On the right, we show the field of the conserved quantity $\psi$ for a zoomed-in region of the system (dashed box in Fig.~\ref{fig:fast_rates}(a)). 
Within this region, we determine the average concentrations in mesoscopic environments (colored spots) that are smaller than the length scale of the traveling waves. Using a standard clustering algorithm~\cite{scikit-learn}, we identify whether there are two distinct phases and their corresponding average concentrations. 
To the left of the figure, we display these phase concentrations in their corresponding colors, together with the equilibrium phase diagram. In these local environments, the concentrations lie on the binodal manifold whenever the wave is in an $A$-rich domain. However, in the domains that are rich in $B$ and $C$, the system is well-mixed and the concentration lies on the plane of constant $\psi$ (light grey).

Identifying local equilibria has become a guiding principle for studying various active systems~\cite{Halatek2018, Halatek2018b, Fritsch2021,Robinson2025}. 
Here, we demonstrate that phase separation can control the spatiotemporal organization of reaction-diffusion systems, where different phase equilibria can coexist simultaneously at different parts of traveling waves.

\begin{acknowledgements}
J.B. thanks the German Research Foundation for financial support through the DFG Project BA 8210/1-1. 
G.B. thanks the Agencia Estatal de Investigación for funding through the Juan de la Cierva postdoctoral programme JDC2023-051554-I. 
A.K. is partly supported by John Templeton Foundation (grant 62828) and by the European Union’s Horizon 2020 research and innovation programme under the Marie Skłodowska-Curie Grant Agreement No. 101068029.    
\end{acknowledgements}

\clearpage
\newpage
\section*{Appendix}
\subsection*{Interaction parameters}
In the Flory-Huggins free energy density, as introduced in Eq.~\eqref{eq:free_ener}, molecular interactions between component $A$ and the other components $j=B,C,S$ are parameterized by $\chi_{Aj}$.
As explained in the main text, we restrict ourselves to cases where $A$ phase separates from the solvent, while $B$ and $C$ only localize into the $A$-rich and $A$-poor phases, but do not build phases themselves.

In a binary mixture between $A$ and $S$, the free energy density has zero curvature at $\rho_A = 0.5$ when $\chi_{AS}=2$.
When increasing $\chi_{AS}$ to higher values, a larger and larger non-convex domain, centered around $\rho_A = 0.5$, exists in the binary mixture. Thus, we have chosen $\chi_{AS}>2$ for our model.
To ensure that $B$ and $C$ do not phase separate from $A$, we choose $\chi_{AB}<2$ and $\chi_{AC}<2$. Here, even a binary mixture of $A$ and $B$, or $ A$ and $ C$, respectively, cannot form phases.

\subsection*{Response to perturbation of phase equilibria}
\label{app:perturb_eq}
Minimizing the free energy density Eq.~\eqref{eq:favg} subject to the constrain~\eqref{eq:avg_con} leads to the conditions of identical chemical potentials $\mu_i (\rr^\text{I/II}) = \partial f (\rr^\text{I/II}) / \partial \rho_i$ and an identical osmotic pressure between the phases~\cite{Safran2019,Kardar2007},
\begin{gather}
    \mu_i(\rrI) = \mu_i(\rrII) \;,  \; \; \;\; \text{for: } i=A,B,C \nonumber \\
     f(\rrI)-f(\rrII) = \sum_{i=A,B,C} \mu_i(\rr^\text{I/II}) (\rho_i^\text{I} - \rho_i^\text{I}) \, .
     \label{eq:phase_eq}
\end{gather}

With these conditions, %
we can find in closed-form the %, in linear order, for the 
linear response of the phase equilibrium to a perturbation of the average concentrations $\rravg$.  %$d\mathcal{E}_n / d\rho_j $.
To do so, we define the  vector $\boldsymbol{\mathcal{E}}=(\mathcal{E}_1,...,\mathcal{E}_7)=(v,\rho^\text{I}_A,\rho^\text{I}_B,\rho^\text{I}_C,\rho^\text{II}_A,\rho^\text{II}_B,\rho^\text{II}_C)$. To impose the constraints~\eqref{eq:phase_eq}, we also introduce the seven-dimensional vector  %We also define the seven-dimensional vector % with the dummy variables $\mathcal{E}_{1,...,7}=v,\rho^\text{I}_A,\rho^\text{I}_B,\rho^\text{I}_C,\rho^\text{II}_A,\rho^\text{II}_B,\rho^\text{II}_C$, and 
$\boldsymbol{\mathcal{C}}$ with the elements $\mathcal{C}_{1,2,3}=v \rho^\text{I}_{A,B,C} + (1-v)\rho^\text{I}_{A,B,C}- \rho_{A,B,C}$ (fixing $\rho_C=\psi-\rho_A-\rho_B$), $\mathcal{C}_{4,5,6}=\mu_{A,B,C}(\rrI) - \mu_{A,B,C}(\rrII)$, and $\mathcal{C}_{7} = f(\rrI) - f(\rrII) - \sum_{i} \mu_i(\boldsymbol{\rho}^\text{I/II}) (\rho_i^\text{I}- \rho_i^\text{II})$.
For every phase equilibrium, $\boldsymbol{\mathcal{C}}=\boldsymbol{0}$. To calculate the response of the phase equilibria, % to a perturbation of the average concentration $\rravg$, 
%To ensure that also the perturbations in the average concentrations are phase equilibria, thus $\boldsymbol{\mathcal{C}}=\boldsymbol{0}$, 
we  find % , in linear order, the response
$d\boldsymbol{\mathcal{E}} / d \rho_{A,B}$ by solving
\begin{equation}
%    \boldsymbol{J} \cdot d \boldsymbol{\mathcal{E}} / d \rho_{A,B} + d \boldsymbol{\mathcal{C}}/d \rho_{A,B} =\boldsymbol{0} \; ,
    \boldsymbol{J} \cdot \partial \boldsymbol{\mathcal{E}} / \partial \rho_{A,B} + \partial \boldsymbol{\mathcal{C}}/\partial \rho_{A,B} =\boldsymbol{0} \; ,    
\end{equation}
where $\boldsymbol{J}$ is the Jacobian with entries $J_{ij}=\partial \mathcal{C}_i/\partial \mathcal{E}_j$.

To perform linear stability analysis of the fixed point, as in Eq.~\eqref{eq:pert0}, we combine the Jacobian $\boldsymbol{J}$ and the values of ${\partial (d \rho_i/dt)}/{\partial \mathcal{E}_n }$.
%, which may be computed from Eq.~\eqref{eq:dynps}.

\subsection*{Comparison between the two dynamics} 
\begin{figure}[t!] 
    \includegraphics[width=0.96\linewidth]{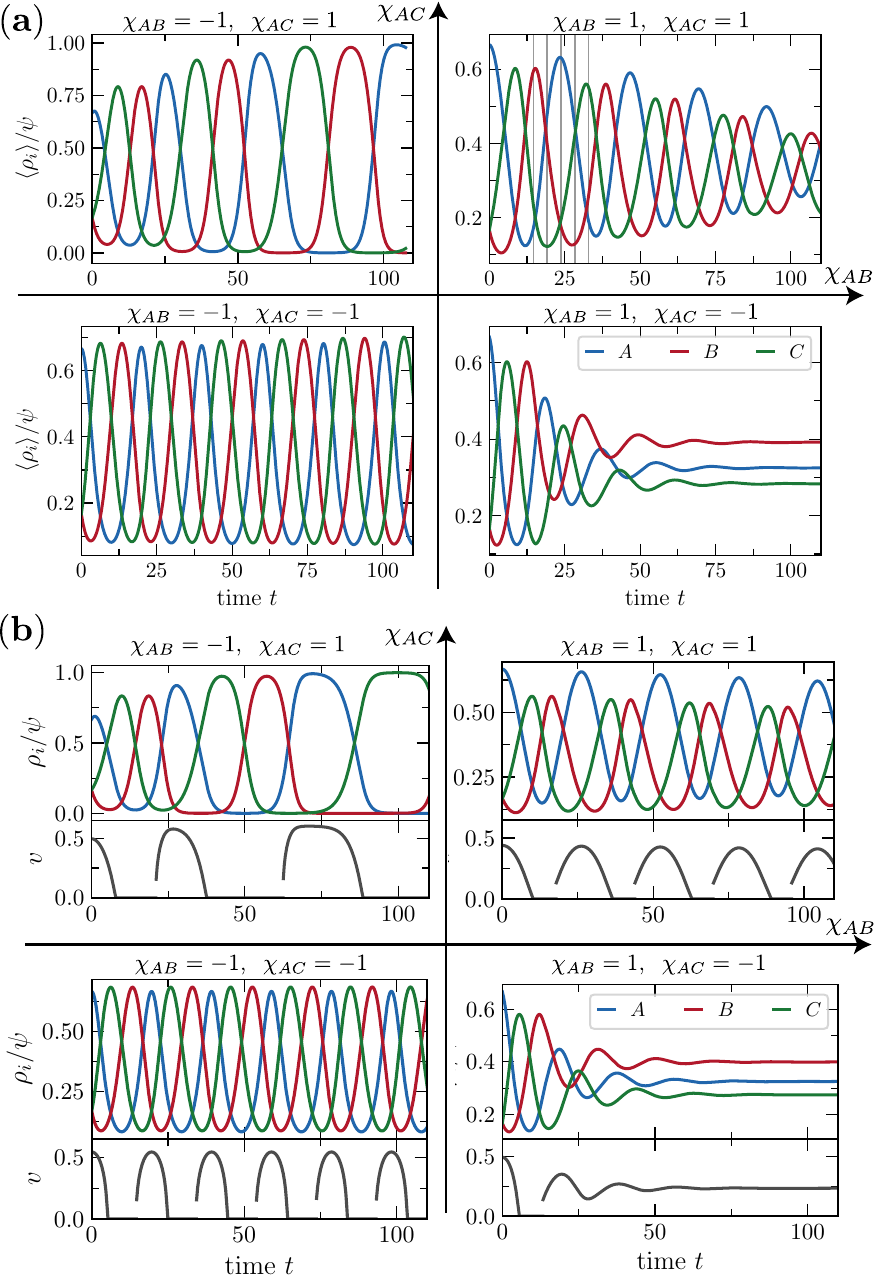}
    \caption{\textbf{Qualitative agreement} between the dynamics at phase equilibrium and the spatial dynamics in the fast diffusion regime:
    \label{fig:spatial_slow} 
    Dynamics of the average concentrations $\rravg$ for four cases in a two-dimensional spatial model (a) and the dynamics at phase equilibrium (b). 
    For the dynamics in (b), we have taken into account the metastability of homogeneous phases, leading to jumps in the relative phase volume $v$.
    The spatial concentration profiles at five time points (grey lines - (a) top right) are shown in Fig.~\ref{fig:phase_space};
    $\psi = 0.6$, $\chi_{AS} = 4$, see numerical details at the end of the Appendix.
    } 
\end{figure}

Here, we compare the dynamics at phase equilibrium with those in spatial models in the limit of fast diffusion.
For this, we show the dynamics of the average concentrations $\langle \rr \rangle=\int_V d\bm{x} \rr(\bm{x})/\int_V d\bm{x}$ of two-dimensional spatial fields and the coarse-grained ordinary differential equation model with metastability, as described in the main text and illustrated in Fig.~\ref{fig:spatial_slow}.
We consider four different systems, corresponding to each quadrant of the $\chi_{AB}-\chi_{AC}$ plane shown in Fig.~\ref{fig:LSA}. 

Qualitatively, both dynamics describe:
For $\chi_{AB}<0,\chi_{AC}>0$ (top left), the oscillation amplitude increases. In every oscillation, initially several droplets or stripes are nucleated and grow, but later they shrink and finally dissolve, such that the system becomes spatially homogeneous again. 
On the contrary, when $\chi_{AB}>0,\chi_{AC}<0$ (bottom right), after the first oscillation where all droplets dissolve, the system stays within the binodal regime and enters a stage where oscillations are suppressed and droplets slowly ripen.
For the two cases $\chi_{AB}=\chi_{AC}>0$ (top right) and $\chi_{AB}=\chi_{AC}<0$ (bottom left), the amplitude of the oscillations slowly decreases/increases. This behavior is a result of the metastability.
Nevertheless, the frequency of the oscillations for $\chi_{AB}=\chi_{AC}<0$ is faster than for $\chi_{AB}=\chi_{AC}>0$.

However, quantitatively, there are differences between the two dynamics. As mentioned in the main text, the full spatial model includes effects from interface regions, the Laplace pressure induced by curved interfaces, finite nucleation time scales even within the spinodal, and ripening of droplets.

% % In the Supplemental Material, we provide movies that show the dynamical evolution of the two-dimensional concentration fields~\cite{sm}. Furthermore, in Fig.~\ref{fig:spatial_slow}(b), we show the spatial profiles of the three components at five time points within one period of the oscillation (grey lines in Fig.~\ref{fig:spatial_slow}(a), top right), illustrating the nucleation, growth, shrinking, and dissolution of droplets.

% In Fig.~\ref{fig:spatial_slow}, we showed how phase separation controls oscillations by using a spatial model. For an easy comparison to our dynamics at phase equilibria, we show in Fig.~\ref{fig:four_cases_ODE} the average concentration trajectories for the corresponding cases.

\subsection*{Numerical details of the spatial models} 
For solving the dynamics of Eq.~\eqref{eq:spatial} for the three components $A$, $B$, and $C$, we use a Pseudo-spectral solver, applying a second-order Runge-Kutta integration scheme.
Every system shown here is a square box with length $L$, discretized by $N$ grid points in each dimension, and subject to periodic boundary conditions. In all of our plots, we report time in units of RPS kinetics (i.e., so that $k_{ij}=1$). However, % plots always report time in terms of RPS to RPS   
%While every timescale in the main part is defined such that reaction rates are $k_{ij}=1$, 
for convenience in the numerical solver, here we vary the timescale by changing the mobility coefficient $\gamma$ together with a length scale of the interface width, set by $\kappa$. 

% \begin{figure}[!t] 
% \includegraphics[width=0.96\linewidth]{./figs/fig8.pdf}
%     \caption{\textbf{Trajectories with dynamics at phase equilibrium} taking the metastability into account:
%      $\psi = 0.6$, $\chi_{AS} = 4$
%     \label{fig:four_cases_ODE} 
%     } 
% \end{figure}

For Fig.~\ref{fig:phase_space}(b) and Fig.~\ref{fig:spatial_slow}(a), we have chosen: 
$\gamma=1$, $L=200$, $N=256$, $\kappa_A=\kappa_B=\kappa_C = 1.5$, $k_{AB}=k_{BC}=k_{CA}=0.0005$, simulated for $T = 250000$.
We initialized the fields homogeneously with $\rho_A = 0.4$, $\rho_B = 0.1$, $\rho_C = 0.1$ and added fluctuations, by shuffling $0.001$ between $A$ and the two other components such that the total concentration is fixed at $\psi = 0.6$.

For Fig.~\ref{fig:fast_rates} and the corresponding movie $\text{SI}\_\text{mov}6$, we have chosen: $\gamma=1$, $L=750$, $N=1024$, $\kappa_A=\kappa_B=\kappa_C = 1.5$, $k_{AB}=k_{BC}=k_{CA}=0.1$, simulated for $T = 10000$.
For the initialization of random fields, we first generate a smooth random field for $\psi$ with local values between 0 and 1 by applying a Gaussian filter in Fourier space ($\sigma = 50$ grid points) to Gaussian noise, then rescaling it to have average total concentration 0.6. From this, we create three fields $\rho_A$, $\rho_B$, and $\rho_C$ using the same Fourier-space filtering method and normalize them so that their sum equals the local $\psi$ at each point.

\clearpage
\end{document}